\documentclass[conference]{IEEEtran}
\IEEEoverridecommandlockouts
\renewcommand{\baselinestretch}{1.1}
\usepackage{cite}
\usepackage{amsmath,amssymb,amsfonts}
\usepackage{algorithmic}
\usepackage{graphicx}
\usepackage{comment}
\usepackage{textcomp}
\usepackage{xcolor}
\usepackage{amsmath,graphicx}
\usepackage{multirow}
\usepackage{graphicx}
\usepackage{booktabs}
\usepackage{comment}
\usepackage{tikz}
\usepackage{bbding}
\usepackage{pifont}
\usepackage{wasysym}
\usepackage{textcase}
\usepackage{url}
\usepackage{amssymb}
\usepackage{siunitx}
\def\BibTeX{{\rm B\kern-.05em{\sc i\kern-.025em b}\kern-.08em
    T\kern-.1667em\lower.7ex\hbox{E}\kern-.125emX}}
\begin{document}

\title{On the Use of Self-Supervised Representation Learning for Speaker Diarization and Separation\\}

\begin{comment}

\author{\textit{Anonymous submission for ASRU 2025}}

\author{\IEEEauthorblockN{Séverin Baroudi}
\IEEEauthorblockA{\textit{Université de Toulon}, \textit{LIS} \\
Toulon, France \\
severin.baroudi@lis-lab.fr}
\and
\IEEEauthorblockN{Hervé Bredin}
\IEEEauthorblockA{\textit{pyannoteAI} \\
Toulouse, France \\
herve@pyannote.ai}
\and
\IEEEauthorblockN{Joseph Razik}
\IEEEauthorblockA{\textit{Université de Toulon}, \textit{LIS}}
Toulon, France \\
joseph.razik@univ-tln.fr
\and
\IEEEauthorblockN{Ricard Marxer}
\IEEEauthorblockA{\textit{Université de Toulon}, \textit{LIS}, \textit{CNRS}}
Toulon, France \\
ricard.marxer@lis-lab.fr
}

\end{comment}

\author{
  Séverin Baroudi$^{1}$,
  Hervé Bredin$^{2}$,
  Joseph Razik$^{1}$,
  Ricard Marxer$^{1}$\\[0.5em]
  $^{1}$Université de Toulon, Aix Marseille Univ, CNRS, LIS, Toulon, France\\
  $^{2}$IRIT / pyannoteAI, Toulouse, France\\[0.5em]
  \texttt{\{severin.baroudi,joseph.razik,ricard.marxer\}@lis-lab.fr},\\
  \texttt{herve@pyannote.ai}
}

\maketitle

\begin{abstract}
Self-supervised speech models such as wav2vec2.0 and WavLM have been shown to significantly improve the performance of many downstream speech tasks, especially in low-resource settings, over the past few years. Despite this, evaluations on tasks such as Speaker Diarization and Speech Separation remain limited. This paper investigates the quality of recent self-supervised speech representations on these two speaker identity-related tasks, highlighting gaps in the current literature that stem from limitations in the existing benchmarks—particularly the lack of diversity in evaluation datasets and variety in downstream systems associated to both diarization and separation.
\end{abstract}


\section{Introduction}
\label{sec:intro}

The self-supervised learning (SSL) paradigm for speech representations has shown great promise in recent years. By improving a broad range of downstream tasks such as Automatic Speech Recognition (ASR) or Speaker Emotion recognition \cite{superb}, they demonstrate great generalization for either transcription or classification speech tasks. While much of the current literature explores the use of SSL representations in the context of ASR, the Speaker Diarization (SD) and Source Separation (SS) tasks have received comparatively less attention. 
Speaker Diarization is the task of partitioning an audio stream into segments according to the identity of active speakers. The purpose of this task is to essentially answer the question of ``who speaks when ?'' in an audio file. 
On the contrary, Speech Separation is the task of estimating the target speech of each speaker from a mixture while also removing background noise and other interferences. The difficulty of this task highly depends on the number of speakers to separate, as well as the acoustic conditions of the recordings. \\

Recent work related to either SD or SS, such as \cite{diarizen,conv_wavlm,ss_ssl} tend to solely leverage one main speech SSL foundation model (WavLM \cite{WavLM}) with the  intent of improving the performance related to either of these tasks. While this approach is valuable for obtaining the best-performing system, it overlooks a deeper evaluation of these internal representations across a broader range of SSL models, such as HuBERT \cite{HuBERT} or wav2vec2.0 \cite{wav2vec2}. Another significant shortcoming of these types of studies is the lack of evaluation regarding key factors related to speech foundation models, such as the type of pretraining dataset and the nature of the upstream pretext task (predictive or contrastive). Finally, although large-scale evaluations—such as those in \cite{large_scale_eval} and \cite{WavLM}—have been carried out using the Speech processing Universal PERformance Benchmark \cite{superb} (SUPERB) across a wide range of speech foundation models for SD and SS, certain limitations remain. One such limitation concerns the dataset (LibriMix \cite{LibriMix})) used by SUPERB to evaluate both tasks, which is derived from the same corpus (LibriSpeech-960h \cite{librispeech}) used for the pretraining of some SSL models such as wav2vec2.0, HuBERT, and WavLM, thus potentially leading to domain mismatch and biased results. Additionally, differences in the downstream architectures of SD and SS, between the models used by SUPERB, and more recent approaches may influence the degree of activation in the internal representations of the SSL model, raising concerns about the validity of the layer-wise analyses conducted across models by these benchmarks. \\

To address these limitations, we make the following key contributions:
\begin{itemize}
    \item We provide a benchmark of the raw performance of a range of commonly used SSL architectures that vary in their pretraining dataset (LibriSpeech, or custom conversational datasets \cite{conv_wavlm}), size and pretext task. Our goal is not to optimize the performance of each system in an attempt to beat the current state-of-the-art, but rather to establish a clear comparison of which architecture is best suited for each task, based solely on the raw representations provided by a given speech foundation model. To ensure this, we focus our study on frozen SSL models rather than fine-tuned versions.
    \item We conduct our evaluation on widely used benchmark datasets for diarization (DIHARD3 \cite{DIHARD}) and separation (WSJ0 \cite{wsj}) respectively, as a mean to mitigate the domain mismatch that we previously introduced, stemming from benchmarking on LibriMix only.
    \item To provide a benchmark on recent SD and SS models, we make use of the neural end-to-end segmentation approach from \textit{Pyannote} \cite{pyannote} for diarization and the time-domain only architecture \cite{TASNET} for separation, instead of the downstream models provided by SUPERB. We evaluate three distinct masking architectures for the latter (ConvTasNet \cite{ConvTasNet}, DPRNN  \cite{DPRNN} and SepFormer \cite{SepFormer}), in order to assess the consistency of the results.
    \item We provide a complete layer-wise analysis of each SSL model for each task, thus allowing us to compare the impact of the pretraining dataset and the pretext task on the downstream evaluation. Furthermore, by leveraging the latest speech foundation model, w2v-BERT 2.0 \cite{W2v-BERT} from MetaAI \footnotemark[1] —which integrates both contrastive and predictive learning methods—we are able to directly assess the impact of these two training paradigms across the evaluated SSL models.
\end{itemize}

\section{Self-Supervised Learning for speech}

Recent self-supervised approaches for speech take their inspiration from Natural Language Processing models (such as BERT \cite{bert}), and possess a common backbone used to learn and produce speech representations. They first encode audio speech into latent representations using a Convolutional Neural Network (CNN) of various kernels and strides. 
Each encoded frame essentially represents \SI{20}{\milli\second} of audio at \SI{16}{\kilo\hertz}, which results in a frame rate of \SI{50}{\hertz}. Random temporal segments of the CNN output are thereafter masked  following a given masking ratio.
\begin{table*}[ht]
  \caption{Summary and evaluation of different self-supervised speech representation models on various TasNet based speech separation architectures (using the WSJ0-2 Mix test set), and on the speaker diarization pipeline (using the DIHARD~3 test set). For separation, the reported metric is the SDRi (dB). As for diarization, FA, MD and SC refer to False Alarm rate, Missed Detection rate and Speaker Confusion rate respectively. Denoising refers to the pretext task of Speech Denoising introduced in the WavLM article \cite{WavLM}.} 
  \label{tab:evaluation}
  \centering
  \resizebox{\linewidth}{!}{
  \begin{tabular}{c | c | c c c | c | c || c | c | c || l  l  l |  c l }
  \cmidrule(lr){1-14}
  \multicolumn{7}{c||}{\textbf{Model Summary}} & \multicolumn{8}{c}{\textbf{Evaluation Results}} \\
  \cmidrule(lr){1-14}
  \multirow{5}{*}{\textbf{Self-Supervised}} & \multirow{5}{*}{\textbf{Pretraining}} & \multicolumn{3}{c|}{\textbf{Learning Task}} & \multirow{5}{*}{\textbf{Layer}}  &  \multirow{5}{*}{\textbf{Model}}  & \multicolumn{3}{c||}{\textbf{Separation}} & \multicolumn{4}{c}{\textbf{Diarization}}  \\
  \cmidrule(lr){3-5} \cmidrule(lr){8-10} \cmidrule(lr){10-14}
    \multirow{-5}{*}{\rotatebox{0}{\textbf{Model}}} & \multirow{-5}{*}{\rotatebox{0}{\textbf{Dataset}}}  & \rotatebox{90}{\textbf{Predictive}} & \rotatebox{90}{\textbf{Contrastive}} & \rotatebox{90}{\textbf{Denoising}} & \multirow{-5}{*}{\rotatebox{0}{\textbf{Size}}} & \multirow{-5}{*}{\rotatebox{0}{\textbf{Size}}} & \multirow{-5}{*}{\parbox[l]{1.55cm}{\textbf{ConvTasNet} \\ \makebox[1.75cm][c]{(SDRi $\uparrow$)}}} & \multirow{-5}{*}{\parbox[l]{1.22cm}{\textbf{DPRNN} \\ \makebox[1.2cm][c]{(SDRi $\uparrow$)}}}  & \multirow{-5}{*}{\parbox[l]{1.32cm}{\textbf{SepFormer} \\ \makebox[1.4cm][c]{(SDRi $\uparrow$)}}} & \multirow{-5}{*}{\rotatebox{0}{\parbox[t]{0.01cm}{FA \\ (\%)}}} & \multirow{-5}{*}{\rotatebox{0}{\parbox[t]{0.01cm}{MD \\ (\%)}}} & \multirow{-5}{*}{\rotatebox{0}{\parbox[t]{0.01cm}{SC \\ (\%)}}} & \multicolumn{0}{c}{\multirow{-5}{*}{\rotatebox{0}{\parbox[c]{0.92cm}{\textbf{DER} $\downarrow$  \makebox[0.9cm][c]{\textbf{(\%)} }}}}} & \\ 
  \cmidrule(lr){1-14}
  wav2vec2.0~\cite{wav2vec2} \footnotemark[1]        & \multirow{2}{*}{Librispeech}  &  & $\checkmark$ &  & \multirow{2}{*}{12} & \multirow{3}{*}{95M} &  17.0  & 19.2 & 20.2 & 4.8  & 7.7 & 7.9 & 20.4 \\
  HuBERT~\cite{HuBERT}  \footnotemark[1]            &          \multirow{2}{*}{(960h)}                          & $\checkmark$ &  &  &        \multirow{2}{*}{Transf.}                    & 95M                     &  17.0  & 18.9 & 20.3  & 4.6 & 8.1 & 7.6 & 20.3 \\
  WavLM   ~\cite{WavLM} \footnotemark[1]      &                                    & $\checkmark$ &  & $\checkmark$ &                            &                     &  17.5  & 19.2 & 20.9  & 5.4 & 7.2 & 8.2 & 20.9  \\
  \cmidrule(lr){1-14}
  Conversational & Conversational & \multirow{2}{*}{$\checkmark$} &  & \multirow{2}{*}{$\checkmark$} & 12 & \multirow{2}{*}{95M} & \multirow{2}{*}{17.8}  & \multirow{2}{*}{19.7} & \multirow{2}{*}{21.2}  & \multirow{2}{*}{5.1} & \multirow{2}{*}{6.4} & \multirow{2}{*}{5.7} & \multirow{2}{*}{17.3}\\
  WavLM \cite{conv_wavlm} & Speech (663h) & & & & Transf. &  & &   & & & & \\
  \cmidrule(lr){1-14}
   \multirow{2}{*}{w2v-BERT \footnotemark[2] \cite{W2v-BERT}}           & Assembled  &  \multirow{2}{*}{$\checkmark$} &  \multirow{2}{*}{$\checkmark$} &  & 24       & \multirow{2}{*}{581M} & \multirow{2}{*}{\textbf{17.9}} & \multirow{2}{*}{\textbf{19.9}} & \multirow{2}{*}{\textbf{21.5}} & \multirow{2}{*}{4.9} & \multirow{2}{*}{6.3} & \multirow{2}{*}{5.1} & \multirow{2}{*}{\textbf{16.4}} \\
   & dataset (4.5Mh) & & & &  Conf. & & & & & & & \\
  \cmidrule(lr){1-14}
  Baseline / Reproduction & - &  & - &  & - & - & 15.6 \cite{ConvTasNet} / 16.4 & 19.0 \cite{DPRNN} / 18.0 &  20.5 \cite{SepFormer} / 20.4 & 6.2 & 8.1 & 7.6 & 22.0 \\
  \cmidrule(lr){1-14}
  \end{tabular}
  }
\end{table*}
The masked features are fed into a succession of Transformer layers (resp. Conformer \cite{conformer} layers for w2v-BERT), for contextual prediction of the masked sections using self-attention based mechanisms. The predicted context vector is compared to target pseudo-labels.\\

The target generation varies by architecture, with HuBERT using discretized hidden units~\cite{HuBERT}, predicted via k-means clustering over the training set, and wav2vec2.0 leveraging quantized codebook units~\cite{vqwav2vec}. The former applies a Masked Language Modeling (MLM) loss, while the latter uses contrastive loss. While HuBERT solely trains on single speaker speech content derived from the audiobook dataset Librispeech, the SSL model WavLM \cite{WavLM} attempts to specialize self-supervised learning with hidden units towards multi-speaker content by applying an utterance mixing strategy to the audio input. %Two utterances of single speaker content from a batch are effectively mixed together (with a main speaker being majoritary), before being fed to the CNN encoder. Since the targets yet represent the clean speech of the main speaker, a denoïsing process is implicitely performed by WavLM to predict the original and majoritary single speaker unit. \\
On the other hand, Conversational WavLM \cite{conv_wavlm} relies on a different pretraining dataset and leverages an assembled multi-domain corpus made of 663 hours of real-world conversational datasets such as DIHARD-3 \cite{DIHARD}, AiSHELL-4 \cite{AISHELL}, or AMI \cite{AMI}. \\

Finally, w2v-BERT \cite{W2v-BERT} benefits from both contrastive and predictive learning to provide meaningful speech representations. The architecture features 24 Conformer layers, in which the first half supports the contrastive aspect of the model, while the second half enhances its predictive component. In a similar fashion as wav2vec2.0, the target context vectors (for the contrastive part) are generated using a quantization module that uses codewords derived from codebooks. This quantizer is also used to generate the associated hidden units to predict, that will serve the predictive part of the model. The last 12 conformer layers of the model follow the same learning strategy as HuBERT, leveraging the MLM loss using the output of the last layer as the context vector and units of the quantizer as the target. The model combines both learning aspects by jointly optimizing the contrastive and MLM losses, and is pretrained on 4.5Mh of audio data from over 143 different languages.

\section{Evaluating Speaker Diarization}

Our baseline for evaluating the speaker diarization task is the latest open source \textit{pyannote 3.1} toolkit \cite{pyannote21}, which has demonstrated state-of-the-art performance on multiple diarization datasets. It features a local supervised speaker segmentation model combined with global unsupervised clustering \cite{cluster1}. The segmentation model features a frame-wise feature extractor (SincNet \cite{sincnet}) that ingests 10s long audio chunks. The latter serves as the input of multiple bi-directional LSTM and linear layers that perform multi-class classification in the \textit{powerset} space \cite{powerset}. The \textit{powerset} considers one class for non-speech, one class for each single speaker, and one for each unique pair of overlapping speakers. 
Once trained, the segmentation model outputs classes of active speakers on 10s long chunks. To process an entire audio file, it is required to stitch each local window. To do so, speaker embeddings are extracted on single-only active speaker segments of each window, and Agglomerative Hierarchical Clustering (AHC) is performed to globally attribute the identity of each speaker from each chunk. Diarization Error Rate (DER) is the metric used for evaluating each of our systems. The latter is computed by summing three specific error rates over an audio chunk: speaker confusion, false alarm and missed detection rates. \\

\begin{comment}

\begin{figure}[t]
  \centering
  \includegraphics[width=\linewidth]{1.png}
  \caption{End-to-End Segmentation model using Self-Supervised representations}
  \label{fig:segmentation}
\end{figure}
\end{comment}
\footnotetext[1]{\url{https://pytorch.org/audio/main/pipelines}}
\footnotetext[2]{\url{https://huggingface.co/facebook/w2v-bert-2.0}}

\begin{figure*}[t]
  \centering
  \includegraphics[width=\linewidth]{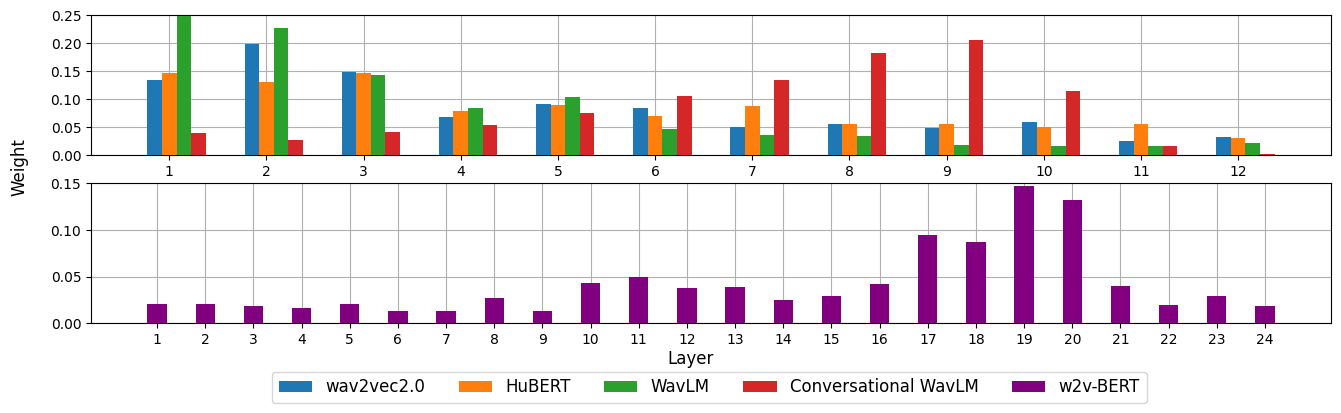}
  \caption{Layer-wise analysis of each self-supervised models for the speaker diarization task.}
  \label{fig:layer-wise_diar}
\end{figure*}

To assess each self-supervised model, the SincNet feature extractor was replaced by a frozen SSL model. We then extract the representations directly from raw 10s audio chunks, along each Transformer (or Conformer) layer and apply learnable weighted average on the representations (such as $\sum \alpha_i = 1$, where $\alpha$ represents the weight related to the $i^{th}$ ``former" layer). These weights (reported in Figure \ref{fig:layer-wise_diar} after training), will serve as a way to understand which part of the model contributes the most towards our task.  For complete evaluation, we train our segmentation models on the diverse and multi-domain DIHARD 3 dataset \cite{DIHARD}. Since this dataset provides no \textit{training} set, we use 77\% of the DIHARD 3 \textit{development}  set, as our training  dataset, and the remaining 23\% as our true development set. The \textit{test} set remains unchanged. In order to get global diarization results, we extract speaker embeddings using pretrained ResNet-34 embedding model from the Wespeaker toolkit \cite{wespeaker}. The segmentation models (2.2M parameters) are trained for approximately 100 epochs on a single V100-32GB. \\

As shown in Table \ref{tab:evaluation}, switching from the SincNet-based feature extractor to off-the-shelf self-supervised representations improves performance by 8\% relative on average, mostly on the false alarm rate. It shows that the SSL representations help the segmentation model at finding when a speaker is active. If we compare the three models pretrained on librispeech, we can see that HuBERT and wav2vec2.0 are almost evenly matched. %Surprisingly, WavLM shows the worst performance out of all the SSL models. This result is contradictory to the one experimentally observed on the DIHARD 3 \textit{development} set (\cite{specializing}), clearly demonstrating that WavLM is the better candidate when compared to its predecessors, due to added denoising pretext task. \\ 
Surprisingly, WavLM offers the worst performance out of all the SSL models, even though the metrics observed during training showed that it outperformed both HuBERT and wav2vec2.0 on the \textit{training} and \textit{development} sets.
%This result is contradictory to the one experimentally observed on the DIHARD 3 \textit{development} set (\cite{specializing}), clearly demonstrating that WavLM is the better candidate when compared to its predecessors, due to added denoising pretext task.
On the contrary, Conversational WavLM offers a huge boost in performance (approximately 20\% relative over models pretrained on librispeech, and 26\% over SincNet), improving all three sub-components of DER. This result shows that specializing a self-supervised model by training it specifically on conversational datasets instead of single speaker content plays a huge part into increasing the performance. Finally, w2v-BERT improves over Conversational WavLM by 5\% relative, and 25\% over off-the-shelf models, by leveraging a bigger architecture (24 Conformer layers), as well as a higher and more diverse amount of unlabeled audio content (4.5Mh) used for pretraining. \\

%, achieving a new state-of-the-art over the previous 16.8\% \cite{ANSD-MA-MSE}. 
By looking at the layer activation weights (from Fig. \ref{fig:layer-wise_diar}), SSL models pretrained on librispeech have their first Transformer-based layers activating the most (mainly layers 1,2 and 3), indicating that the early part of the model contributes the most towards the speaker diarization task. While wav2vec2.0 and HuBERT show a similar activation score, WavLM demonstrates an even earlier and higher activation than them, mostly towards layers 1 and 2. 

On the contrary, Conversational WavLM mainly activates mid-to-top layers of the architecture (layers 7,8,9 and 10). Furthermore, the activation seems more homogeneous than for other models. Having layers that contribute the most towards diarization being located towards the top part of the model appears to have a significant impact on the performance, as Conversational WavLM offers better results over previous models that activated early layers only. It is also an indication that the model is making better use of the representations coming from the whole network. This assumption is further verified by the activation scores observed on w2v-BERT.
On the latter, two activation spikes are observed, respectively associated with the contrastive (layers 10,11 and 12), and the predictive (layers 17,18,19 and 20) top parts of the model, indicating that w2v-BERT tries to leverage both types of pretraining learning tasks to solve diarization. 
According to the results observed on this last SSL model, a predictive oriented learning task appears to offer a better contribution than a contrastive one. Finally, differences in the SSL pretraining dataset plays a significant role in the layer activations observed, as the librispeech models, when compared to Conversational WavLM or w2v-BERT, show a very different layer-wise activation for speaker diarization.

\begin{comment}
\begin{table*}[th]
  \caption{Evaluation of different self-supervised speech representation models on various TasNet based speech separation architectures (using the WSJ0-2 Mix test set), and on the speaker diarization pipeline (using the DIHARD~3 test set). For separation, the reported metric is the SDRi (dB). As for diarization, FA, MD and SC refer to False Alarm rate, Missed Detection rate and Speaker Confusion rate respectively. } 
  \label{tab:evaluation}
  \centering
  \resizebox{\linewidth}{!}{
  
  \begin{tabular}{l | c c || c  c  c | c | }
  \toprule
  Feature extraction & ConvTasNet & Dual-Path RNN & FA\% & MD\% & SC\% & DER\% \\
  \midrule
  Vanilla Model        &  15.6 \cite{ConvTasNet} / 16.4  & 19.0 \cite{DPRNN} / 18.0 & 6.2 & 8.1 & 7.6 & 21.6 \cite{specializing} \\
  \midrule
  wav2vec2.0~\cite{wav2vec2}         & 17.0  & 19.2 & 4.8  & 7.7 & 7.9 & 20.4 \\
  HuBERT~\cite{HuBERT}             & 17.0  & 18.9 & 4.6  & 8.1 & 7.6 & 20.3 \\
  WavLM~\cite{WavLM}              & 17.5  & 19.2 & 5.4  & 7.2 & 8.2 & 20.9  \\
  \midrule
  Conversational WavLM \cite{specializing}  & 17.8  & 19.7 & 5.1  & 6.4 & 5.7 & 17.3\\
  \midrule
  w2v-BERT 2.0~\cite{w2v-BERT}        & 17.9 & 19.9 & 4.9 & 6.3 & 5.1 & 16.4 \\

  \bottomrule
  \end{tabular}
  }
\end{table*}

\begin{figure*}[t]
  \centering
  \includegraphics[width=\linewidth]{8.png}
  \caption{Layer-wise analysis of the tested self-supervised models for the speaker separation task (ConvTasNet used as the masking network).}
  \label{fig:layer-wise_sep}
\end{figure*}

\end{comment}

\section{Evaluating Speaker Separation}

\begin{figure}[t]
  \centering
  \includegraphics[width=\linewidth]{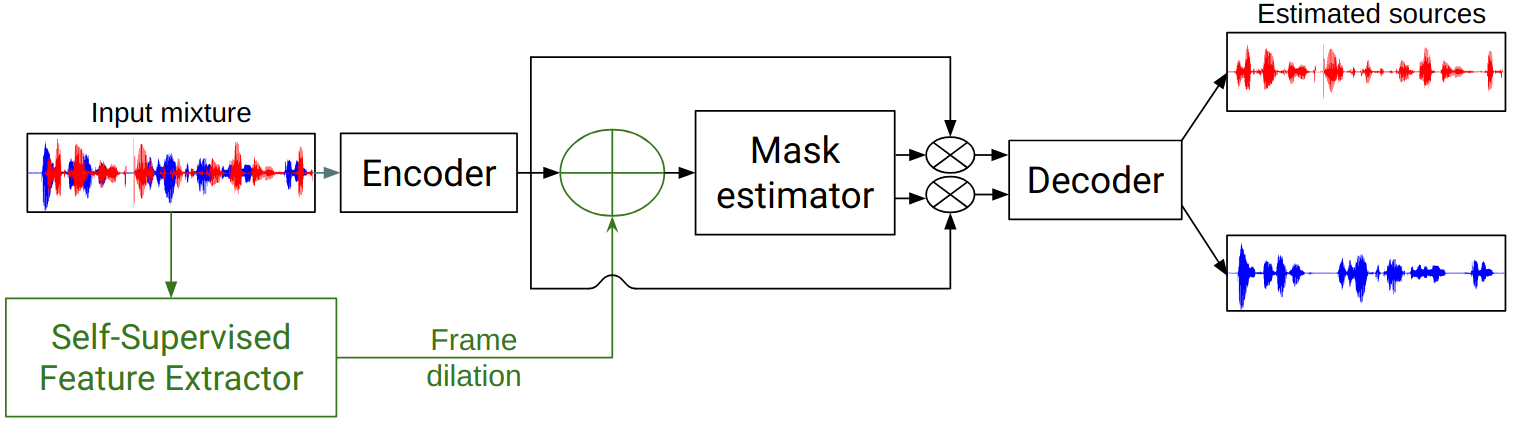}
  \caption{TasNet architecture using SSL Representations}
  \label{fig:tasnet}
\end{figure}

The baseline for evaluating Speaker Separation is the Time-domain audio separation Network \cite{TASNET} (TasNet). Previous frameworks relied on short-time Fourier Transform (STFT) to perform separation, but several limitations were encountered when attempting to process both the magnitude and phase associated with the mixture \cite{phase1}. As a result, time-domain only neural networks using an encoder-decoder based architecture proved to be more suited for speech separation. The TasNet architecture features an encoder (in the form of a Conv1D module) applied directly on the mixture. The resulting latent space is then processed through a masking network, before being decoded (using transposed Conv1D), as displayed in Figure \ref{fig:tasnet}. \\

%Knowing the number of sources to estimate, this block generates an equal number of corresponding masks, which are vectors with values ranging between 0 and 1.
%The masks are subsequently multiplied (element-wise) by the latent features of the encoder, generating an estimate for each source. The latter is then processed through the decoder which is a transposed Conv1D layer. Using a specific learnable encoder-decoder system (instead of a STFT framework) allows the model to pass the mixture into a latent space that is more suited for masking computation, as well as waveform reconstruction. 

Three masking networks are evaluated in this study, to assess whether different masking strategies could provide consistent results among each SSL architecture. The first one is the ConvTasNet \cite{ConvTasNet}, which features multiple layers of depthwise (applied along the frames) and pointwise (applied along the features) convolutions of varying kernel and stride (used to capture long-range dependencies on the input mixture).
\begin{comment}
The second evaluated network is the DPRNN \cite{DPRNN}. It first segments the latent space $y$ (containing $N$ features per frame) into chunks of a fixed size $K$ (with a 50\% overlap between chunks) that are then concatenated to form a 3D tensor of size $N$x$K$x$S$ (with $S$ representing the number of chunks). The latter is then subsequently processed multiple times along the $K$ and $S$ axis using LSTM layers. The idea is to process both locally and globally the features and the frames provided by the latent space $y$. In order to generate as many masks as needed, a $Conv2D$ layer is used to extend the size of the 3D tensor (along the feature dimension) to $N$x$n$. Finally, it is reshaped to reconstruct one output $y$ per speaker using $overlap-add$. 
\end{comment}
The second one is the DPRNN \cite{DPRNN}, which first segments the latent representations of the encoder into chunks of a fixed size (with a 50\% overlap between chunks) that are then concatenated to form a 3D tensor of size. It is then subsequently processed multiple times along the chunk and frame axes using LSTM layers, to process the features and the frames provided by the encoder both locally and globally. In order to generate as many masks as needed, a $Conv2D$ layer is used to extend the size of the 3D tensor (along the feature dimension) to the number of sources to predict. Finally, it is reshaped to reconstruct one output per speaker using $overlap-add$ before being decoded. Lastly, SepFormer \cite{SepFormer} improves upon DPRNN by introducing Transformer-based modules in place of LSTMs, leveraging attention mechanisms to extend the context window across the frame and chunk dimensions of the 3D tensor.

\begin{figure*}[t]
  \centering
  \includegraphics[width=\linewidth]{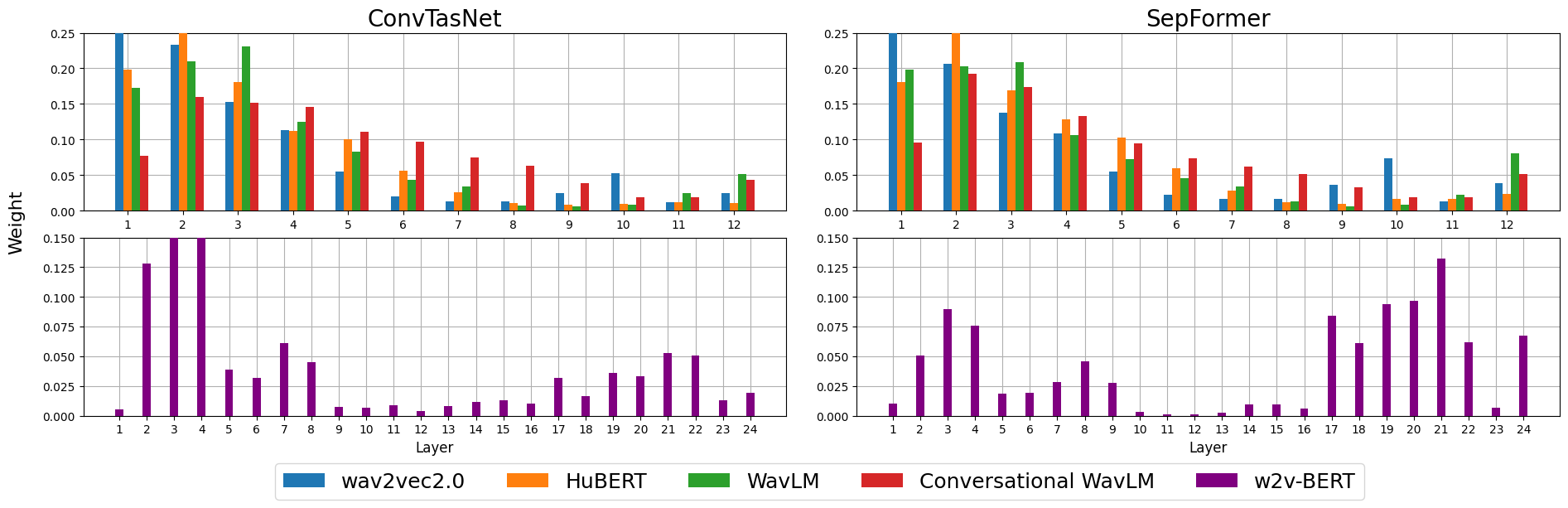}
  \caption{Layer-wise analysis of each self-supervised models for the speaker separation task, using two masking networks : ConvTasNet and SepFormer. (DPRNN exhibited a trend similar to that of ConvTasNet; therefore, we decided not to display it).}
  \label{fig:layer-wise_sep}
\end{figure*}

To evaluate the performance of the speaker separation task, Source-to-Distortion Ratio Improvement (SDRi) is used as a metric to quantify the fidelity of the reconstructed waveform over the original signal. The latter is calculated by substracting the SDRs of both the estimated source and the mixture. SDR is calculated by computing the logarithmic scale of the ratio between the clean signal and the estimated distortions.
The distortions (used to compute SDR) are inferred as the residual error between the true signal and its corresponding estimated one.
\subsection{Implementing Self-Supervised Representations}

Since the TasNet network makes use of an encoder-decoder based architecture, the SSL models cannot be used as a direct substitute of the encoder, because we do not possess an effective way to decode SSL representations to reconstruct temporal waveforms.
%Since the TasNet model makes use of an encoder-decoder based architecture, the SSL representations extracted from the input mixture cannot be directly used as features of the masking network.
Rather, in our setup, SSL representations are concatenated onto the latent representations of the encoder, and serve as additional information, as depicted in Figure \ref{fig:tasnet}.

Usually, TasNet models perform better at the highest encoding framerate \cite{DPRNN} (meaning small kernel and stride sizes). Conversely, SSL feature extractors produce representations at the fixed framerate $f_{SSL}$ of \SI{50}{\hertz}. To match the framerates of both modules and properly align them, we replicate the SSL features along the frames to obtain the same number of frames as the encoder.
For this task, we evaluate each model on the WSJ0-2mix \cite{wsj} \SI{8}{\kilo\hertz} dataset. Since the tested self-supervised models are pretrained on \SI{16}{\kilo\hertz} audio content, we upsample the audio mixture beforehand (from 8 to \SI{16}{\kilo\hertz}), using Kaiser-windowed sinc interpolation. Once the sources are estimated, we downsample them back to \SI{8}{\kilo\hertz} for loss computation. 

To generate the SSL representations, we apply the same extraction strategy as for diarization (using learnable weighted average on the layers). 
ConvTasNet (5.1M parameters), DPRNN (14.8M parameters) and SepFormer (25.9M parameters) are trained using the Asteroid \cite{asteroid} and Speechbrain \cite{Speechbrain} libraries for 200, 50 and 100 epochs, respectively, using the provided recipes. We applied learning rate halving with a patience of 3 epochs, starting after 30 epochs of training. Each experiment is conducted on a single A40-48GB.

\begin{comment}
\begin{figure}[t]
  \centering
  \includegraphics[width=\linewidth]{20.png}
  \caption{TasNet architecture using Self-Supervised Speech Representations}
  \label{fig:tasnet}
\end{figure}
\end{comment}

\subsection{Benchmark Results}

Final results are reported in Table \ref{tab:evaluation} in the Separation section. For the baseline section (without using SSL representations), scores shown on the left side represent the SDRi reported by the authors of each paper, while the right side are scores we achieved after training ourselves with the basic recipe provided by each library. \\

Results show that integrating SSL representations into the masking network consistently improves the separation capability of the models. For ConvTasNet, we notice a minimum 0.6 dB (3.7\% relative) increase in performance when using wav2vec2.0 or HuBERT. WavLM improves the SDRi by a larger margin of 0.9 dB (5.5\% relative) over the vanilla model. Finally, the best results were achieved when using Conversational WavLM and w2v-BERT, as both models managed to improve the SDRi by 1.4 or 1.5 dB (9\% relative) respectively. 
For DPRNN, larger gains can be observed, as using SSL features from off-the-shelf models improve the SDRi by 7\% relative over the baseline. Interestingely, WavLM and wav2vec2.0 offered similar performance (19.2 dB), as opposed to HuBERT which slightly underperformed. Conversational WavLM and w2v-BERT both increase the gap with off-the-shelf model, like with the ConvTasNet (9\% and 11\% relative increase in performance, when compared to the baseline). However, for the SepFormer architecture, a clear bottleneck emerges when integrating features from wav2vec2.0 and HuBERT. Specifically, their addition leads to degraded performance compared to the vanilla SepFormer. Furthermore, the gains from using more suited representations such as WavLM or Conversational WavLM are notably smaller than those observed with other separation networks. For instance, the improvement in SDRi with WavLM is limited to only 0.5 dB over the baseline, whereas other models exhibit gains exceeding 1 dB. This suggests that the information encoded in these representations may be redundant or superfluous for a more sophisticated masking-based separation network.\\

Finally, by looking at the layer-wise analysis for the task of speech separation (Fig. \ref{fig:layer-wise_sep}), we observe a similar behavior as for the speaker diarization task when looking at the off-the-shelf models only: the most activated layers are located around the early parts of the model. However, the same comparison cannot be made for both Conversational WavLM and w2v-BERT, as we see a similar activation as the ones observed with librispeech pretrained SSL models. When looking at Conversational WavLM, which showed that layers 7,8,9 and 10 were mostly active for diarization, we can see layers 2,3,4 and 5 being used the most for speech separation. \\
 
A similar behavior can be seen with w2v-BERT. Since both speaker diarization and speech separation are speaker identity related tasks, one would assume that the activated layers is similar between the two tasks, but this does not appear to be the case for models pretrained on conversational datasets. A hypothesis could be that the latter are more sensitized towards speaker-turns (and predicting parts of a masked conversation), which play an important part in diarization. On the other hand, the speaker separation task processes speech mixtures only, which relies more on the low-level acoustic features of the speech rather than the longer-term patterns of a conversation. \\

While the layer-wise analysis reveals consistent trends across the various masking networks considered in this study, we observe that, for w2v-BERT, the predictive layers exhibit higher activation than the contrastive ones when comparing SepFormer with ConvTasNet. This suggests that representations learned through a predictive self-supervised learning task may align more effectively with SepFormer than those obtained via contrastive learning.

\section{CONCLUSION}

In this article, we investigated the use of self-supervised speech
representations for the tasks of speaker diarization and speech separation. We first review the state-of-the-art regarding SSL architectures associated with speech (wav2vec2.0, HuBERT, WavLM,w2v-BERT and Conversational WavLM)), as well as the downstream models associated to each task (EEND segmentation model for diarization, and TasNet for separation), before conducting a performance study over a selection of the most prominent self-supervised models. Supplemented by a layer-wise analysis, we found that the pretraining dataset, the learning task, and the type of downstream model, significantly impacts the performance. Although Conversational WavLM and w2v-BERT hugely differ in the amount of hours of speech used for pretraining (663h vs 4.5Mh), as well as in the number of parameters (95M vs 581M), they offer comparable results for both tasks, emphasizing the importance of the type of data used rather than its quantity. The analysis also revealed a performance degradation in certain SSL models, for speech separation, when combined with more effective masking networks. Finally, leveraging DIHARD3 for diarization and WSJ0 for separation allowed us to effectively evaluate model performance free from any dataset bias, allowing us to gain insight into the transferability of the observed performance gain to more recent architectures \cite{SepFormer,mossformer}. \\

%Finally, we conduct a layer-wise analysis of each SSL model to assess their individual contributions to each task. This approach aims to provide insight into how each model performs in relation to diarization or separation. While assumptions regarding certain models were verified (such as off-the-shelf SSL models activating early layers for speaker identity related tasks), interesting results were observed regarding the activation layers of models pretrained on conversational datasets (Conversational WavLM and w2v-BERT 2.0), depending on the task. 
To further investigate the discrepancy in the layer contribution observed with conversational SSL models for speaker diarization, as opposed to speech separation, it would be interesting to identify the type of information (phonetic or semantic) that the layers can generate, and oppose them to off-the-shelf models. One hypothesis would be that the representations extracted  from an off-the-shelf model (such as WavLM) and a conversational one (like Conversational WavLM or w2v-BERT) are complementary. An other point of interest is the feature fusion technique employed for the separation task. While feature concatenation with the TasNet encoder is our mainly used technique for the evaluation of this specific task, more sophisticated and performing approaches involving cross attention or Feature-wise Linear Modulation (FiLM) \cite{film} should be considered for future work, as they could potentially lead to different results and behaviors from the layers of the SSL model. Finally, the layer-wise analyses presented in this work open up interesting possibilities for applying these insights to joint models that integrate both diarization and separation (such as \cite{PixIT}).

\section{Acknowledgments}

This work was granted access to the HPC resources of IDRIS
under the allocations AD011014044R2 and 2024-AD011015163 made by GENCI, and
supported by the Agence de l’Innovation Defense under the
grant number 2022 65 0079. It also benefited from the support of the French National Research Agency through the ANR-20-CE23-0012-01 (MIM) grant.

\bibliographystyle{IEEEtran}
\bibliography{refs.bib}

@article{WavLM,
author = {Chen, Sanyuan and Wang, Chengyi and Chen, Zhengyang and Wu, Yu and Liu, Shujie and Chen, Zhuo and Li, Jinyu and Kanda, Naoyuki and Yoshioka, Takuya and Xiao, Xiong and Wu, Jian and Zhou, Long and Ren, Shuo and Qian, Yanmin and Qian, Yao and Zeng, Michael and Yu, Xiangzhan and Wei, Furu},
year = {2022},
month = {10},
pages = {1-14},
title = {WavLM: Large-Scale Self-Supervised Pre-Training for Full Stack Speech Processing},
volume = {16},
journal = {IEEE Journal of Selected Topics in Signal Processing},
doi = {10.1109/JSTSP.2022.3188113}
}

@inproceedings{conv_wavlm,
	title        = {Specializing Self-Supervised Speech Representations for Speaker Segmentation},
	author       = {Baroudi, Séverin and Pellegrini, Thomas and Bredin, Hervé},
	year         = 2024,
	month        = {09},
	pages        = {3769--3773},
        booktitle    =  {25th Annual Conference of the International Speech Communication Association, Interspeech 2024}
}

@ARTICLE{HuBERT,

  author={Hsu, Wei-Ning and Bolte, Benjamin and Tsai, Yao-Hung Hubert and Lakhotia, Kushal and Salakhutdinov, Ruslan and Mohamed, Abdelrahman},

  journal={IEEE/ACM Transactions on Audio, Speech, and Language Processing}, 

  title={HuBERT: Self-Supervised Speech Representation Learning by Masked Prediction of Hidden Units}, 

  year={2021},

  volume={29},

  number={},

  pages={3451-3460},

  doi={10.1109/TASLP.2021.3122291}}

@inproceedings{superb,
  author={Shu-wen Yang and Po-Han Chi and Yung-Sung Chuang and Cheng-I Jeff Lai and Kushal Lakhotia and Yist Y. Lin and Andy T. Liu and Jiatong Shi and Xuankai Chang and Guan-Ting Lin and Tzu-Hsien Huang and Wei-Cheng Tseng and Ko-tik Lee and Da-Rong Liu and Zili Huang and Shuyan Dong and Shang-Wen Li and Shinji Watanabe and Abdelrahman Mohamed and Hung-yi Lee},
  title={{SUPERB: Speech Processing Universal PERformance Benchmark}},
  year=2021,
  booktitle={Proc. Interspeech 2021},
  pages={1194--1198},
  doi={10.21437/Interspeech.2021-1775}
}

@inproceedings{librispeech,
	location = {South Brisbane, Queensland, Australia},
	title = {Librispeech: An {ASR} corpus based on public domain audio books},
	isbn = {978-1-4673-6997-8},
	url = {http://ieeexplore.ieee.org/document/7178964/},
	doi = {10.1109/ICASSP.2015.7178964},
	shorttitle = {Librispeech},
	eventtitle = {{ICASSP} 2015 - 2015 {IEEE} International Conference on Acoustics, Speech and Signal Processing ({ICASSP})},
	pages = {5206--5210},
	booktitle = {2015 {IEEE} International Conference on Acoustics, Speech and Signal Processing ({ICASSP})},
	publisher = {{IEEE}},
	author = {Panayotov, Vassil and Chen, Guoguo and Povey, Daniel and Khudanpur, Sanjeev},
	urldate = {2023-08-03},
	date = {2015-04},
	langid = {english},
}

@article{AMI,
  author = {Carletta, Jean},
  title = {Announcing the AMI Meeting Corpus},
  journal = {The ELRA Newsletter},
  volume = {11},
  number = {1},
  pages = {3--5},
  year = {2006},
  url = {http://groups.inf.ed.ac.uk/ami/corpus/overview.shtml}
}

@inproceedings{AISHELL,
  title     = {AISHELL-4: An Open Source Dataset for Speech Enhancement, Separation, Recognition and Speaker Diarization in Conference Scenario},
  author    = {Yihui Fu and Luyao Cheng and Shubo Lv and Yukai Jv and Yuxiang Kong and Zhuo Chen and Yanxin Hu and Lei Xie and Jian Wu and Hui Bu and Xin Xu and Jun Du and Jingdong Chen},
  year      = {2021},
  booktitle = {Interspeech 2021},
  pages     = {3665--3669},
  doi       = {10.21437/Interspeech.2021-1397},
  issn      = {2958-1796},
}

@inproceedings{pyannote,
  author={Hervé Bredin},
  title={{pyannote.audio 2.1 speaker diarization pipeline: principle, benchmark, and recipe}},
  year=2023,
  booktitle={Proc. INTERSPEECH 2023},
  pages={1983--1987},
  doi={10.21437/Interspeech.2023-105}
}

@INPROCEEDINGS{sincnet,

  author={Ravanelli, Mirco and Bengio, Yoshua},

  booktitle={2018 IEEE Spoken Language Technology Workshop (SLT)}, 

  title={Speaker Recognition from Raw Waveform with SincNet}, 

  year={2018},

  volume={},

  number={},

  pages={1021-1028},

  doi={10.1109/SLT.2018.8639585}}

@inproceedings{powerset,
  author={Alexis Plaquet and Hervé Bredin},
  title={{Powerset multi-class cross entropy loss for neural speaker diarization}},
  year=2023,
  booktitle={Proc. INTERSPEECH 2023},
  pages={3222--3226},
  doi={10.21437/Interspeech.2023-205}
}

@INPROCEEDINGS{wespeaker,

  author={Wang, Hongji and Liang, Chengdong and Wang, Shuai and Chen, Zhengyang and Zhang, Binbin and Xiang, Xu and Deng, Yanlei and Qian, Yanmin},

  booktitle={ICASSP 2023 - 2023 IEEE International Conference on Acoustics, Speech and Signal Processing (ICASSP)}, 

  title={Wespeaker: A Research and Production Oriented Speaker Embedding Learning Toolkit}, 

  year={2023},

  volume={},

  number={},

  pages={1-5},

  doi={10.1109/ICASSP49357.2023.10096626}}

@inproceedings{wav2vec2,
author = {Baevski, Alexei and Zhou, Henry and Mohamed, Abdelrahman and Auli, Michael},
title = {wav2vec 2.0: a framework for self-supervised learning of speech representations},
year = {2020},
isbn = {9781713829546},
publisher = {Curran Associates Inc.},
address = {Red Hook, NY, USA},
booktitle = {Proceedings of the 34th International Conference on Neural Information Processing Systems},
articleno = {1044},
numpages = {12},
location = {Vancouver, BC, Canada},
series = {NIPS'20}
}

@inproceedings{PixIT,
  title     = {PixIT: Joint Training of Speaker Diarization and Speech Separation from Real-world Multi-speaker Recordings},
  author    = {Joonas Kalda and Clément Pagés and Ricard Marxer and Tanel Alumäe and Hervé Bredin},
  year      = {2024},
  booktitle = {The Speaker and Language Recognition Workshop (Odyssey 2024)},
  pages     = {115--122},
  doi       = {10.21437/odyssey.2024-17},
}

@inproceedings{film,
  author    = {Ethan Perez and Florian Strub and Harm de Vries and Vincent Dumoulin and Aaron Courville},
  title     = {FiLM: Visual Reasoning with a General Conditioning Layer},
  booktitle = {Proceedings of the Thirty-Second AAAI Conference on Artificial Intelligence (AAAI-18)},
  year      = {2018},
  pages     = {3942--3951},
  publisher = {AAAI Press},
  url       = {https://ojs.aaai.org/index.php/AAAI/article/view/11671},
  doi       = {10.1609/aaai.v32i1.11671}
}

@INPROCEEDINGS{mossformer,
  author={Zhao, Shengkui and Ma, Bin},
  booktitle={ICASSP 2023 - 2023 IEEE International Conference on Acoustics, Speech and Signal Processing (ICASSP)}, 
  title={MossFormer: Pushing the Performance Limit of Monaural Speech Separation using Gated Single-head Transformer with Convolution-augmented Joint Self-Attentions}, 
  year={2023},
  }

@unpublished{LibriMix,
  TITLE = {{LibriMix: An open-source dataset for generalizable speech separation}},
  AUTHOR = {Cosentino, Joris and Pariente, Manuel and Cornell, Samuele and Deleforge, Antoine and Vincent, Emmanuel},
  URL = {https://inria.hal.science/hal-03354695},
  NOTE = {working paper or preprint},
  YEAR = {2020},
  MONTH = May,
  HAL_ID = {hal-03354695},
  HAL_VERSION = {v1},
}

@inproceedings{DIHARD,
  title     = {The Third DIHARD Diarization Challenge},
  author    = {Neville Ryant and Prachi Singh and Venkat Krishnamohan and Rajat Varma and Kenneth Church and Christopher Cieri and Jun Du and Sriram Ganapathy and Mark Liberman},
  year      = {2021},
  booktitle = {Interspeech 2021},
  pages     = {3570--3574},
  doi       = {10.21437/Interspeech.2021-1208},
  issn      = {2958-1796},
}

@ARTICLE{large_scale_eval,
  author={Yang, Shu-wen and Chang, Heng-Jui and Huang, Zili and Liu, Andy T. and Lai, Cheng-I and Wu, Haibin and Shi, Jiatong and Chang, Xuankai and Tsai, Hsiang-Sheng and Huang, Wen-Chin and Feng, Tzu-hsun and Chi, Po-Han and Lin, Yist Y. and Chuang, Yung-Sung and Huang, Tzu-Hsien and Tseng, Wei-Cheng and Lakhotia, Kushal and Li, Shang-Wen and Mohamed, Abdelrahman and Watanabe, Shinji and Lee, Hung-yi},
  journal={IEEE/ACM Transactions on Audio, Speech, and Language Processing}, 
  title={A Large-Scale Evaluation of Speech Foundation Models}, 
  year={2024},
  volume={32},
  number={},
  pages={2884-2899},
  doi={10.1109/TASLP.2024.3389631}}

@INPROCEEDINGS{ss_ssl,
  author={Chen, Zhuo and Kanda, Naoyuki and Wu, Jian and Wu, Yu and Wang, Xiaofei and Yoshioka, Takuya and Li, Jinyu and Sivasankaran, Sunit and Eskimez, Sefik Emre},
  booktitle={ICASSP 2023 - 2023 IEEE International Conference on Acoustics, Speech and Signal Processing (ICASSP)}, 
  title={Speech Separation with Large-Scale Self-Supervised Learning}, 
  year={2023},
  volume={},
  number={},
  pages={1-5},
  doi={10.1109/ICASSP49357.2023.10096876}}

@article{W2v-BERT,
  author       = {Yu{-}An Chung and
                  Yu Zhang and
                  Wei Han and
                  Chung{-}Cheng Chiu and
                  James Qin and
                  Ruoming Pang and
                  Yonghui Wu},
  title        = {W2v-BERT: Combining Contrastive Learning and Masked Language Modeling
                  for Self-Supervised Speech Pre-Training},
  journal      = {CoRR},
  volume       = {abs/2108.06209},
  year         = {2021},
  url          = {https://arxiv.org/abs/2108.06209},
  eprinttype    = {arXiv},
  eprint       = {2108.06209},
  bibsource    = {dblp computer science bibliography, https://dblp.org}
}

@article{ConvTasNet,
  author       = {Yi Luo and
                  Nima Mesgarani},
  title        = {TasNet: Surpassing Ideal Time-Frequency Masking for Speech Separation},
  journal      = {CoRR},
  volume       = {abs/1809.07454},
  year         = {2018},
  url          = {http://arxiv.org/abs/1809.07454},
  eprinttype    = {arXiv},
  eprint       = {1809.07454},
  bibsource    = {dblp computer science bibliography, https://dblp.org}
}

@article{DPRNN,
  title={Dual-Path RNN: Efficient Long Sequence Modeling for Time-Domain Single-Channel Speech Separation},
  author={Yi Luo and Zhuo Chen and Takuya Yoshioka},
  journal={ICASSP 2020 - 2020 IEEE International Conference on Acoustics, Speech and Signal Processing (ICASSP)},
  year={2019},
  pages={46-50},
  url={https://api.semanticscholar.org/CorpusID:204575939}
}

@INPROCEEDINGS{SepFormer,
  author={Subakan, Cem and Ravanelli, Mirco and Cornell, Samuele and Bronzi, Mirko and Zhong, Jianyuan},
  booktitle={ICASSP 2021 - 2021 IEEE International Conference on Acoustics, Speech and Signal Processing (ICASSP)}, 
  title={Attention Is All You Need In Speech Separation}, 
  year={2021},
  volume={},
  number={},
  pages={21-25},
  doi={10.1109/ICASSP39728.2021.9413901}}

@INPROCEEDINGS{TASNET,
  author={Luo, Yi and Mesgarani, Nima},
  booktitle={2018 IEEE International Conference on Acoustics, Speech and Signal Processing (ICASSP)}, 
  title={TaSNet: Time-Domain Audio Separation Network for Real-Time, Single-Channel Speech Separation}, 
  year={2018},
  volume={},
  number={},
  pages={696-700},
  doi={10.1109/ICASSP.2018.8462116}}

@inproceedings{bert,
    title = "{BERT}: Pre-training of Deep Bidirectional Transformers for Language Understanding",
    author = "Devlin, Jacob  and
      Chang, Ming-Wei  and
      Lee, Kenton  and
      Toutanova, Kristina",
    editor = "Burstein, Jill  and
      Doran, Christy  and
      Solorio, Thamar",
    booktitle = "Proceedings of the 2019 Conference of the North {A}merican Chapter of the Association for Computational Linguistics: Human Language Technologies, Volume 1 (Long and Short Papers)",
    month = jun,
    year = "2019",
    address = "Minneapolis, Minnesota",
    publisher = "Association for Computational Linguistics",
    url = "https://aclanthology.org/N19-1423/",
    doi = "10.18653/v1/N19-1423",
    pages = "4171--4186"
}

@inproceedings{conformer,
  title     = {Conformer: Convolution-augmented Transformer for Speech Recognition},
  author    = {Anmol Gulati and James Qin and Chung-Cheng Chiu and Niki Parmar and Yu Zhang and Jiahui Yu and Wei Han and Shibo Wang and Zhengdong Zhang and Yonghui Wu and Ruoming Pang},
  year      = {2020},
  booktitle = {Interspeech 2020},
  pages     = {5036--5040},
  doi       = {10.21437/Interspeech.2020-3015},
  issn      = {2958-1796},
}

@inproceedings{vqwav2vec,
  title={vq-wav2vec: Self-Supervised Learning of Discrete Speech Representations},
  author={Baevski, Alexei and Schneider, Steffen and Auli, Michael},
  booktitle={Proceedings of the 8th International Conference on Learning Representations (ICLR)},
  year={2020},
  url={https://openreview.net/forum?id=rylwJxrYDS}
}

@inproceedings{pyannote21,
  author={Hervé Bredin},
  title={{pyannote.audio 2.1 speaker diarization pipeline: principle, benchmark, and recipe}},
  year=2023,
  booktitle={Proc. INTERSPEECH 2023},
  pages={1983--1987},
  doi={10.21437/Interspeech.2023-105},
  issn={2958-1796}
}

@inproceedings{cluster1,
  title     = {Advances in Integration of End-to-End Neural and Clustering-Based Diarization for Real Conversational Speech},
  author    = {Keisuke Kinoshita and Marc Delcroix and Naohiro Tawara},
  year      = {2021},
  booktitle = {Interspeech 2021},
  pages     = {3565--3569},
  doi       = {10.21437/Interspeech.2021-1004},
  issn      = {2958-1796},
}

@INPROCEEDINGS{phase1,
  author={Erdogan, Hakan and Hershey, John R. and Watanabe, Shinji and Le Roux, Jonathan},
  booktitle={2015 IEEE International Conference on Acoustics, Speech and Signal Processing (ICASSP)}, 
  title={Phase-sensitive and recognition-boosted speech separation using deep recurrent neural networks}, 
  year={2015},
  volume={},
  number={},
  pages={708-712},
  doi={10.1109/ICASSP.2015.7178061}}

@INPROCEEDINGS{wsj,
  author={Hershey, John R. and Chen, Zhuo and Le Roux, Jonathan and Watanabe, Shinji},
  booktitle={2016 IEEE International Conference on Acoustics, Speech and Signal Processing (ICASSP)}, 
  title={Deep clustering: Discriminative embeddings for segmentation and separation}, 
  year={2016},
  volume={},
  number={},
  pages={31-35},
  doi={10.1109/ICASSP.2016.7471631}}

@inproceedings{asteroid,
  title     = {Asteroid: The PyTorch-Based Audio Source Separation Toolkit for Researchers},
  author    = {Manuel Pariente and Samuele Cornell and Joris Cosentino and Sunit Sivasankaran and Efthymios Tzinis and Jens Heitkaemper and Michel Olvera and Fabian-Robert Stöter and Mathieu Hu and Juan M. Martín-Doñas and David Ditter and Ariel Frank and Antoine Deleforge and Emmanuel Vincent},
  year      = {2020},
  booktitle = {Interspeech 2020},
  pages     = {2637--2641},
  doi       = {10.21437/Interspeech.2020-1673},
  issn      = {2958-1796},
}

@article{specializing,
  title={Specializing Self-Supervised Speech Representations for Speaker Segmentation},
  author={Baroudi, S{\'e}verin and Pellegrini, Thomas and Bredin, Herv{\'e}}
}

@article{SpeechBrain,
  author  = {Mirco Ravanelli and Titouan Parcollet and Adel Moumen and Sylvain de Langen and Cem Subakan and Peter Plantinga and Yingzhi Wang and Pooneh Mousavi and Luca Della Libera and Artem Ploujnikov and Francesco Paissan and Davide Borra and Salah Zaiem and Zeyu Zhao and Shucong Zhang and Georgios Karakasidis and Sung-Lin Yeh and Pierre Champion and Aku Rouhe and Rudolf Braun and Florian Mai and Juan Zuluaga-Gomez and Seyed Mahed Mousavi and Andreas Nautsch and Ha Nguyen and Xuechen Liu and Sangeet Sagar and Jarod Duret and Salima Mdhaffar and Ga{{\"e}}lle Laperri{{\`e}}re and Mickael Rouvier and Renato De Mori and Yannick Est{{\`e}}ve},
  title   = {Open-Source Conversational AI with SpeechBrain 1.0},
  journal = {Journal of Machine Learning Research},
  year    = {2024},
  volume  = {25},
  number  = {333},
  pages   = {1--11},
  url     = {http://jmlr.org/papers/v25/24-0991.html}
}

@inproceedings{diarizen,
  title={Leveraging self-supervised learning for speaker diarization},
  author={Han, Jiangyu and Landini, Federico and Rohdin, Johan and Silnova, Anna and Diez, Mireia and Burget, Luk{\'a}{\v{s}}},
  booktitle={Proc. ICASSP},
  year={2025}
}

\end{document}